\documentstyle[prd,aps,preprint]{revtex}
        		
\newcommand\ee{\end{equation}}
\newcommand\be{\begin{equation}}
\newcommand\eea{\end{eqnarray}}
\newcommand\bea{\begin{eqnarray}}

\newcommand\GeV{\,\mbox{GeV}}

\newcommand\Mpl{M_{\rm Pl}}

\newcommand\lsim{\mathrel{\rlap{\lower4pt\hbox{\hskip1pt$\sim$}}
    \raise1pt\hbox{$<$}}}
\newcommand\gsim{\mathrel{\rlap{\lower4pt\hbox{\hskip1pt$\sim$}}
    \raise1pt\hbox{$>$}}}

\def\dslash{\not{\hbox{\kern-2pt $\partial$}}}
\def\Dslash{\not{\hbox{\kern-4pt $D$}}}
\def\Oslash{\not{\hbox{\kern-4pt $O$}}}
\def\Qslash{\not{\hbox{\kern-4pt $Q$}}}
\def\pslash{\not{\hbox{\kern-2.3pt $p$}}}
\def\kslash{\not{\hbox{\kern-2.3pt $k$}}}
\def\qslash{\not{\hbox{\kern-2.3pt $q$}}}
 \newtoks\slashfraction
 \slashfraction={.13}
 \def\slash#1{\setbox0\hbox{$ #1 $}
 \setbox0\hbox to \the\slashfraction\wd0{\hss \box0}/\box0 }
 
\def\eeq{\end{equation}}
\def\beq{\begin{equation}}

\begin{document}

\preprint{FERMILAB-PUB-97/236-A}
\draft
\tighten

\title{Comments on $D$-term inflation}
\author{David H.  Lyth}
\address{School of Physics and Chemistry, \\
University of Lancaster, Lancaster LA1 4YB U.~K.}
\author{Antonio Riotto}
\address{NASA/Fermilab Astrophysics Center,\\ Fermilab
National Accelerator Laboratory, 
Batavia, Illinois~~60510.
}
\date{June 1997}
\maketitle

\begin{abstract}
An inflationary stage dominated by a $D$-term avoids the slow-roll problem of inflation in supergravity and can naturally emerge in theories with a non-anomalous or anomalous $U(1)$ gauge symmetry. In  the latter  case, however,  the scale of inflation as imposed by  the COBE  normalization is in contrast with the value  fixed by the Green-Schwarz mechanism of anomaly cancellation. In this paper we discuss      possible solutions to this problem and  comment about the  fact that the string-loop generated Fayet-Iliopoulos $D$-term may trigger  the presence of global and local cosmic strings at the end of inflation.  

\end{abstract}
\newpage
\baselineskip=20pt
{\bf 1.}~~ It is commonly accepted that  inflation \cite{abook}  looks  more natural in supersymmetric theories rather in non-supersymmetric ones. This is because  the necessity of introducing very small
parameters to ensure the extreme flatness of the inflaton potential seems very
unnatural and fine-tuned in  most non-supersymmetric theories, while this naturalness is achieved in supersymmetric models. The nonrenormalization theorems in exact global supersymmetry guarantee that we can fine-tune any parameter at the tree-level and this fine-tuning will not be destabilized by radiative corrections at any order in perturbation theory \cite{grisaru79}. This is the advantage of invoking supersymmetry. There is, however, a severe problem one has to face when dealing with inflation model building in the context of supersymmetric theories. The  generalization of supersymmetry from a global to 
a local symmetry automatically incorporates gravity and, therefore, inflation model building must be considered in the framework of supergravity theories. 

In small-field models of inflation (values of fields smaller than the reduced Planck scale $\Mpl\simeq 2.4\times 10^{18}$ GeV), where the theory is under control,
it is reasonable to work in the context of 
supergravity. This is a relatively recent activity because although 
small-field models were the first to be proposed \cite{new,singlet} 
they were soon abandoned in favour of models with fields first of order
the Planck scale \cite{primordial} and then much bigger \cite{chaotic}.
Activity began again after hybrid inflation was proposed
\cite{l90,LIN2SC}, with the realization
\cite{CLLSW} that the model is again of the small-field type.
In Ref.~\cite{CLLSW} supersymmetric implementations of 
hybrid inflation were considered, in the context of both global 
supersymmetry and of supergravity.

The supergravity potential is rather involved, but it  can still  be written as 
a $D$-term plus an $F$-term, and it is usually supposed that
the $D$-term vanishes during inflation. 
Now, for models where the $D$-term vanishes,
the slow-roll  parameter $\eta=\Mpl^2 V''/V$ generically receives
various contributions of order $\pm 1$. This is the so-called $\eta$-problem of supergravity theories.  This  
crucial point was first emphasized in Ref.~\cite{CLLSW}, though it
is essentially 
a special case of the more general result,
noted much earlier
\cite{dinefisch,coughlan}, that 
there are contributions of order $\pm H^2$ to the mass-squared of every
scalar field. Indeed, in a small-field
model the troublesome contributions to $\eta$ may
be regarded as contributions
to the coefficient $m^2$ in the expansion  of the inflaton
potential. Therefore, it is very difficult to naturally implement a slow-roll inflation in the context of supergravity. The problem basically  arises since inflation, by definition, breaks global supersymmetry because of a nonvanishing cosmological constant (the false vacuum energy density of the inflaton). In supergravity theories, supersymmetry breaking is transmitted by gravity interactions and the squared mass of the inflaton becomes naturally of order of $V/\Mpl^2\sim  H^2$. The perturbative renormalization of the K\"ahler potential is therefore crucial   for the inflationary dynamics due to a non-zero energy
density which breaks supersymmetry spontaneously during inflation. How severe the problem is depends on the magnitude of $\eta$.
If $\eta$ is 
not too small then its smallness could be due to accidental 
cancellations. Having $\eta$ not too small requires 
that the spectral index  $n=1-6\epsilon+2\eta$ ($\epsilon=\frac{1}{2}\Mpl^2(V'/V)^2$ is another slow-roll parameter) 
be not too small, so the observational bound
$|n-1|<0.3$ is already beginning to make an accident look unlikely.

{\bf 2.}~~Several proposals to solve   the $\eta$-problem already exist in the  literature \cite{lr}. One of the most promising solutions is  certainly $D$-term inflation \cite{ewansgrav,bindvali,halyo}.  It is based on the observation that $\eta$ gets contributions of order 1  only if inflation proceeds along a $D$-flat direction or, in other words, when the vacuum energy density is dominated by an $F$-term. On the contrary, if  the  vacuum energy density is dominated by nonzero
$D$-terms  and supersymmetry breaking is of the $D$-type, scalars get supersymmetry soft breaking masses which depend only on their gauge charges. Scalars charged under  
 the corresponding gauge symmetry obtain a mass much larger than $H$, while gauge singlet fields can only get masses from loop  gauge interactions.  In particular, if the inflaton field is identified with a gauge singlet, its potential may be flat up to loop corrections and supergravity corrections to $\eta$ from the $F$-terms are not present. 

If the theory  contains an abelian  $U(1)$  gauge symmetry (anomalous or not), 
the Fayet-Iliopoulos
$D$-term term 
\begin{equation}
\xi \int d^4 \theta ~V = \xi D
\end{equation} 
is gauge invariant and therefore allowed by the symmetries.  It may lead  to $D$-type  supersymmetry breaking. It is important to notice that this term may be present in the underlying theory from the very beginning or appears in the effective theory after some heavy degrees of freedom have been integrated out.
It looks  particularly intriguing, however, 
that an  anomalous $U(1)$
symmetry  is  usually present in string theories \cite{u(1)A}.   The corresponding Fayet-Iliopoulos term is  \cite{fi} 
\begin{equation}
\xi = \frac{g^2}{192\pi^2}\:{\rm Tr} {\bf Q}\:\Mpl^2,
\end{equation}
 where ${\rm Tr} {\bf Q}\neq 0$ indicates the trace over the $U(1)$ charges of the fields present in the spectrum of the theory.
The $U(1)$ group may be assumed to emerge from string theories so that 
and the  anomaly is cancelled by the Green-Schwarz
mechanism. In such a case $\sqrt{\xi}$ is expected to be of the order of the 
stringy scale,   $(10^{17}-10^{18})$ GeV or so.

Let us remind the reader how $D$-term inflation proceeds \cite{bindvali,halyo}. To exemplify the description, let us take the  toy model containing  three chiral superfields $S$, $\Phi_+$ and
$\Phi_-$ with charges equal to $0$, $+ 1$ and $- 1$ respectively under the $U(1)$ gauge symmetry.
The superpotential has the form 
\beq
W = \lambda S\Phi_+\Phi_-.
\end{equation}
The scalar potential in the global supersymmetry limit reads
\begin{equation}
V = \lambda^2 |S|^2 \left(|\phi_-|^2 + |\phi_+|^2 \right) +
\lambda^2|\phi_+\phi_-|^2 + 
{g^2 \over 2} \left(|\phi_+|^2 - |\phi_-|^2  + \xi \right)^2
\end{equation}
where $\phi_{\pm}$ are the scalr fields of the supermultplets $\Phi_{\pm}$,  $g$ is the gauge coupling and $\xi>0$ is a Fayet-Iliopoulos $D$-term. 
The global minimum is supersymmetry conserving, but the gauge group $U(1)$ is spontaneously broken
\begin{equation}
\langle S \rangle  = \langle \phi_+  \rangle = 0, ~~~ \langle \phi_-\rangle  = \sqrt{\xi}.
\end{equation}
However, if we minimize the potential, for  fixed values of $S$, with respect to
other fields, we find that for  $S > S_c = {g \over \lambda}
\sqrt{\xi}$, the minimum is at $\phi_+ =\phi_- = 0$. Thus, for
$S > S_c$ and $\phi_+ =\phi_- = 0$ the tree level potential
has a vanishing curvature in the $S$ direction and large positive
curvature in the remaining two directions $
m_{\pm}^2 = \lambda^2|S|^2 \pm g^2\xi$
For arbitrarily large $S$ the tree level value of the potential remains
constant $V = {g^2 \over 2}\xi^2$ and the  $S$ plays the role of the   inflaton. As stated above, the charged fields get very large masses due to the $D$-term supersymmetry breaking, whereas the gauge singlet field is massless at the tree-level. 

What is crucial is that,  along  the inflationary
trajectory $\phi_{\pm}=0$, $S\gg S_c$, all the $F$-terms vanish and large supergravity corrections to the $\eta$-parameter do not appear. Therefore , we do not need to make any assumption about the structure of the K\"ahler potential for the $S$-field: minimal $S^*S$ and non-minimal quartic terms in the K\"ahler potential $(S^*S)^2$ (or even higher orders) do not contribute in the curvature, since $F_S$ is vanishing
during inflation. 

Since the energy density  is dominated by the
$D$-term, supersymmetry is broken and this amounts to  splitting  the masses of the scalar and fermionic components of $\Phi_{\pm}$. Such splitting results in the
one-loop effective potential for the inflaton field 
\begin{equation}
V_{{\rm 1-loop}} = {g^2 \over 2}\xi^2 \left( 1 + {g^2 \over 16\pi^2} {\rm ln}
{\lambda^2 |S|^2 \over Q^2}\right).
\end{equation}
The end of inflation is determined either by
the failure of the slow-roll conditions or when $S$ approaches $S_c$. 
COBE imposes the following normalization
\begin{equation}
5.3\times 10^{-4} = \frac{V^{3/2}}{V'\Mpl^3}
\end{equation}
which can be written in the equivalent form
\begin{equation}
 \frac{V^{1/4}}{\epsilon^{1/4}}=8\times 10^{16}\:{\rm GeV}. 
\end{equation}
More or less independently of the value of $|S|$ at the end of 
inflation, this gives with the above potential 
\beq
\sqrt{\xi}=6.6\times 10^{15}\: {\rm GeV}.
\end{equation}
Notice that his normalization is independent from the gauge coupling constant $g$, but depends on the numerical coefficient in the one-loop potential. This, in turn, depends upon the particle content of the specific model under consideration. 

The spectral index results
\beq
n=1-\frac{2}{N}=(0.96-0.98).
\end{equation}
Now, the value of $\xi$ looks  too small to be consistent with the value arising in many compactifications of the heterotic string \cite{fi} (even though some level of flexibility is allowed in M-theory \cite{mtheory}). 

 The stabilisation of the dilaton field is also an unsolved problem in heterotic string theories.
One may ask whether the  value of $\sqrt{\xi}$ required by density perturbations can be motivated by a  realistic string theory.
At this point uncertainties come from the fact that
 $\xi$ is always treated  as constant (up to a coarse-graining scale dependence through the gauge-coupling). This is certainly justified in the effective
field theory approach in which $\xi$ is treated as an input parameter.
In string theories the
gauge and gravitational coupling constants are set through the expectation value of the dilaton field  and  the Fayet-Iliopoulos $D$-term actually is a function of the real part of the dilaton field.  Since the dilaton potential most likely is strongly influenced 
by the inflationary dynamics, the  actual value of $\xi$ at the moment
 when observationally interesting scales crossed the horizon during inflation  might be quite different from the  one "observed" today. 
It seems that  entire question is related to  the problem  of the  dilaton
stabilization and it is hard to make any definite statement without knowing the details of the dilaton dynamics during inflation. All the  estimates made above are valid
within an effective field
theory description, in which the gauge and gravitational constants can be treated
as parameters whose inflaton-dependence arises from the coarse-graining
scale-dependence \cite{dvaliriotto,matsuda}.

{\bf 3.}~~One might  ask if the COBE normalization permits a bigger $\xi$, if the 
slope of the potential is altered. Before looking at a couple of 
specific possibilities, we note the general point that a dramatic increase
will {\it not} be possible, unless $g$ is very small. The reason
is that slow-roll inflation requires $\epsilon\ll 1$, so that the COBE 
normalization requires 
$V_0^{1/4}\ll 8\times 10^{16}$\,GeV. In fact, barring a cancellation,
the observational result $|n-1|\lsim 0.2$ requires $6\epsilon
\lsim 0.2$, corresponding to $V^{1/4}\lsim 3.4\times 10^{16}\GeV$.
This translates to
\begin{equation}
\xi^{1/2} \lsim \frac{4.1\times 10^{16}}{g^{1/2}}\:{\,\rm GeV}, 
\end{equation}
but $g^2/4\pi\sim 0.1$ would correspond to $g\sim 1$,
and to increase $\xi^{1/2}$ much above $4\times 10^{16}\GeV$ requires
an unreasonably small $g^2$.

Still the extra order of magnitude is worth having, so let us see how it 
might be done. One possibility is to increase the numerical coefficient
$c$ 
in front of the one-loop term. If we rewrite the one-loop effective potential as $V_{{\rm 1-loop}}\sim V_0\left(1+ c g^2\ln |S|\right)$,  $c=(1/8\pi^2)$ in the example above. This coefficient depends upon the number of degrees of freedom  which are coupled to the field playing the role of the inflaton and is expected to be quite large, especially if the theory is embedded in some grand unified gauge group. For instance,  the superfields $\Phi_{\pm}$ might be interepreted as 126 and $\overline{126}$ Higgs superfields of $SO(10)$ and there might be more the one vector-like pair coupled to the $S$-field.

It is easy to show that the COBE normalization gives $\xi^{1/2}\propto
c^{1/4}$, so a reasonable value $8\pi^2 c\sim 100$ would give us
a modest increase to $\xi^{1/2}\simeq 2\times 10^{16}$\,GeV.

Another possibility is to suppose that the $V'$ is dominated by a 
tree-level term. The simplest possibility is mass term,
$V_S\equiv \frac12 \widetilde m^2\phi^2$ where $\phi\equiv
\sqrt2|S|$ is the inflaton field. This term may be easily generated. It might be present in the theory under the form of supersymmetric mass.   Otherwise,  imagine that  in some sector of the underlying theory supersymmetry is broken by some $F$-term.  Supersymmetry breaking may be communicated to the $S$-field by gravitational interactions\footnote{In this case the fields $\phi_{\pm}$ get also the same mass $\widetilde{m}$, but it is easy to show that, in the regime we will be working, its presence doesn't affect the main conclusions.}
 and in this case $\widetilde{m}^2\sim F^2/\Mpl^2$. Another possibility is that supersymmetry breaking is transmitted   by gauge interactions (with gauge group $ G$). In this   case, the sector which breaks supersymmetry is assumed to first trasmit supersymmetry breaking to some fields (usually called the messangers) charged under  $G$ by gauge interactions. In turn, these messangers are coupled to the field $S$ which receives a nonvanishing soft supersymmetry breaking  squared mass term at two-loops,  $\widetilde{m}^2\sim\alpha_G^2 F$, where $\alpha_G$ is the  gauge coupling constant. 

If $\phi\equiv \sqrt{2}|S|$ is the inflaton field, the potential during 
inflation is
\be
V=V_0 + V_S + V_{{\rm 1-loop}}, 
\ee
where 
\bea
V_0&\equiv& \frac{g^2}{2}\xi^2, \\
V_S &\equiv& \frac12 \widetilde m^2 \phi^2, \\
V_{{\rm 1-loop}}&\equiv& (8\pi^2 c) \frac{g^4\xi^2}{16\pi^2}\ln\left(\frac{\lambda\phi}{
\sqrt2 Q}\right).
\eea
The derivatives are
\bea
V'_S &=& \widetilde m^2 \phi, \\
V'_{{\rm 1-loop}} &=& (8\pi^2 c) \frac{g^4\xi^2}{16\pi^2} \frac1\phi. 
\eea
We want to suppose that $V'_{{\rm 1-loop}}\ll V'_S$ in the interval of interest,
which is
\be
2\frac{g^2}{\lambda^2}\xi < \phi^2 < 2\frac{g^2\xi^2}{\lambda^2}\:{\rm e}^{2x}, 
\label{b1}
\ee
where the lower end corresponds to $\phi_c^2$  and $x$ is defined as  $x=N(n-1)/2$,  $N$ being   the number of $e$-folds after the
COBE scale leaves the horizon. 
If this is satisfied we shall also have $V''_{{\rm 1-loop}}\ll V''_S$.
Also, the loop contribution at the end of inflation
 will then be
\be
\frac{V_{{\rm 1-loop}}}{V_0} \simeq(8\pi c^2) \frac{g^2}{8\pi^2}
\ee
which is much smaller than 1 for any reasonable choice of parameters.
Thus, if Eq.~(\ref{b1}) is satisfied, the loop is negligible in all 
respects.

We also want to assume that $V_0$ 
dominates in the interval of interest, because otherwise
the COBE normalization will need  $\phi\gsim \Mpl$ 
making it unreasonable to assume a constant $\widetilde m$.
This requires
\be
\frac{\widetilde m\phi}{g\xi} \ll 1.
\label{b2}
\ee
Then the COBE normalization can 
be written
\be
\frac{\xi^{1/2}}{\Mpl} = 3\times 10^{-4} (n-1) \frac {{\rm e}^x}{\lambda}. 
\ee
 We used the result $n-1=2\eta$,
thus ignoring $\epsilon$.

One can verify that Eqs.~(\ref{b1}) and (\ref{b2}), with the COBE 
condition, can be written
\be
\xi^{1/2}_{\rm max}(8\pi^2 c) ^{1/2}
{\rm e}^{-x/2} \ll \xi^{1/2} \ll \xi^{1/2}_{\rm max}, 
\ee
where
\bea
\xi^{1/2}_{\rm max} &=& \ll 2.3\times 10^{-2} g^{-1/2}(n-1)^{1/4}
\Mpl\\
&=& (3.6\times 10^{16}{\,\rm GeV})\left(
\frac{n-1}{2} \right)^{1/4} g^{-1/2}.
\eea
The lower limit is indeed lower than the upper limit for reasonable
parameter. 
This confirms the general conclusion, that the combination $(g\xi)^{1/2}
$ cannot be bigger than a few times $10^{16}\GeV$.
Note that as the upper limit is approached, $\epsilon$ becomes 
significant which complicates matters, but barring fine tuning this 
will not change the general conclusion.

Another  possible solution to the mismatching between the value suggested by string theories and the one imposed by the COBE normalization is to completely decouple the origin of $\xi$ from string theories and to envisage that      the   $D$-term is generated  in some  low-energy effective theory after some degrees of freedom have been integrated out. However, to do so, one has presumably to break supersymmetry by some $F$-terms present in the sector which the heavy fields  belong to and to generate the $D$-term by loop corrections. As a result,  it turns out that    $\langle D\rangle\ll \langle F^2\rangle$, unless some fine-tuning is called for, and large supergravity corrections to $\eta$ appear again. Let us give an example. Consider the following superpotential where a $U(1)$ symmetry has been imposed
\begin{equation}
W=\lambda X \left({\bar \Phi}_1 \Phi_1 - m^2 \right)
+M_1 {\bar \Phi}_1 \Phi_2 + M_2 {\bar \Phi}_2 \Phi_1 ~.
\label{orafeq}
\end{equation}
 For $\lambda^2 m^2 \ll M_1^2,M_2^2$, the vacuum of this model is such
that $\langle \phi_i \rangle =\langle {\bar \phi}_i \rangle =0$
($i=1,2$), where  $\bar{\phi}_i$ and $\phi_i$ are the scalar components of the superfields $\bar{\Phi}_i$ and $\Phi_i$,  respectively. Supersymmetry is broken and $F_X=-\lambda^2 m^2$. This means that in the potential a term like $V=(F_X{\bar \phi}_1 \phi_1+{\rm h.c.})$ will appear. It is easy to show that, integrating out the $\phi_{i}$ and $\bar{\phi}_i$ scalar fields, induces a a nonvanishing Fayet-Iliopoulos $D$-term 
\begin{equation}
\xi\simeq \frac{F_X^2}{16\pi^2(M_1^2-M_2^2)}\ln\left(\frac{M_2^2}{M_1^2}\right),
\end{equation}
which is, however,  smaller than $F_X$ and inflation, if any, is presumably dominated by the $F$-term.   

{\bf 4.}~~Another point we would like to comment on is the following: when the field $\phi_{-}$ rolls down to its present day value $\langle\phi_{-}\rangle=\sqrt{\xi}$  to terminate inflation, cosmic strings may be form since the abelian gauge group $U(1)$ is broken to unity \cite{j2}. As it is known, stable cosmic strings arise when the manifold ${\cal M}$ of degenerate vacua has a non-trivial first homotopy group, $\Pi_1({\cal M})\neq {\bf 1}$. 
The fact that at the end of hybrid inflationary models the formation of cosmic strings may occur was already noticed in Ref. \cite{j1} in the context of global supersymmetric theories and in Ref. \cite{linderiotto} in the context of supergravity theories. In $D$-term inflation the string per-unit-length is given by $\mu=2\pi\xi$. Cosmic strings forming at the end of $D$-term inflation are very heavy and temperature anisotropies may arise both from the inflationary dynamics and from the presence of cosmic strings. From recent numerical simulations on the cosmic microwave background  anisotropies induced by cosmic strings \cite{a1,a2}
it is possible to infer than 
this mixed-perturbation scenario \cite{linderiotto}
leads to the COBE normalized value $\sqrt{\xi}=4.7\times 10^{15}$ GeV \cite{j2}, which is of course smaller  than the value obtained in the absence of cosmic strings.  Moreover, cosmic strings contribute to the angular spectrum an amount of order of 75\% in $D$-term inflation \cite{j2}, which might render the angular spectrum, when both cosmic strings and inflation contributions are summed up, too smooth to be in agreement with present day observations  \cite{a1,a2}. 

All these considerations and, above all, the fact that the value of   $\sqrt{\xi}$ is further reduced with respect to the case in which cosmic strings are not present,  would appear  to   exacerbate the problem of reconciling the value of  $\sqrt{\xi}$ suggested by COBE with the value inspired by string theories when cosmic strings are present. However,  even though cosmic strings are generally produced,  this is not always  true. If there is an anomalous $U(1)$ factor in the four-dimensional gauge group, since the Yukawa couplings respect the anomalous $U(1)$,  this becomes global, the local symmetry being broken by the mass of the gauge boson. The global $U(1)$ can  be spontaneously broken by the Fayet-Iliopoulos $D$-term. 
This can be understood in the following way: the  Fayet-Iliopoulos $D$-term  depends upon the value of the dilaton and the gauge anomalous $U(1)$ is always broken through
the dilaton vacuum expectation value.  Indeed in the effective field theory the $U(1)$ symmetry  is realised
nonlinearly from the very
beginning. So whenever effective field theory makes sense the anomalous $U(1)$ gauge  symmetry is  already broken and gauge bosons are massive. More formally, the  
relevant couplings of the dilaton superfield
$s$ to the $U(1)$ gauge superfield  $V$ (with  gauge
invariant field strengths  $W^{\alpha}$) read in the  global limit
\begin{eqnarray}
{\cal L} &=& - \int d^4 \theta \ln \left(s+ s^\dagger -
\delta_{{\rm GS}}V\right) \nonumber \\
&  + &  \int d^2\theta \left[{s \over 4} k {\rm Tr} W^{\alpha} W_{\alpha} +
{\rm h.c.}\right] \label{eq:L} 
\end{eqnarray}
where $\delta_{{\rm GS}}$ is the Green-Schwarz coefficient and 
$k$ is the Kac-Moody level of the group $U(1)$.
Under a $U(1)$ gauge transformation $A_\mu \rightarrow A_\mu
+ \partial_\mu \alpha$,
$s$ is shifted as 
\begin{equation}
s\rightarrow s + {i \over 2} \delta_{{\rm GS}} \alpha(x). \label{eq:shift}
\end{equation}
The gauge boson gets a mass in string theory eating the model-independent axion and 
the residual anomalous global $U(1)$ may be   spontaneously broken by the Fayet-Iliopoulos $D$-term, in which case 
 global cosmic strings are formed. Moreover, in realistic four-dimensional string models, there are extra local $U(1)$ symmetries that can be also spontaneously broken by the $D$-term. This happens necessarily if there are no singlet fields charged under the anomalous $U(1)$ only. In such a case, besides the previously mentioned global cosmic strings, there may arise local cosmic strings associated to the breaking of extra $U(1)$ factors. However,  the condition to produce cosmic strings is $\Pi_1({\cal M})\neq {\bf 1}$ and one must consider the structure of the {\it whole} potential, {\it i.e.} all the $F$-terms and all the $D$-terms. When this is done, it turns out that, depending on the specific models, some or all of the (global and local) cosmic strings may disappear. In general there can be models with anomalous $U(1)$ that have just global strings, just local strings, both global and local strings or, more important,   no strings at all \cite{casas}. The latter case is, to our opinion, the most prefer
able case since the presence of cosmic strings renders the  problem of reconciling the COBE normalized low value of $\xi$ with the one suggested by string theory even worse.

In the case in which the Fayet-Iliopoulos $D$-term is present in the theory from the very beginning  because of  anomaly-free $U(1)$ symmetry and not due to  some underlying string theory, the value $\sqrt{\xi}\sim 10^{15}$ GeV is very natural and is not in conflict with the presence of cosmic strings. The only shortcoming seems to be a  too smooth angular spectrum because cosmic strings may provide most contribution to the angular spectrum. If this problem is taken seriously and one wants to avoid the presence of cosmic strings, 
a natural solution to it is to assume that the  $U(1)$ gauge group is broken before the onset of inflation  so that no cosmic strings will be produced whn $\phi_{-}$ rolls down to its ground state. This may be easily achieved by introducing a pair of vector-like (under $U(1)$) fields 
$\Psi$ and $\bar{\Psi}$ and two gauge singlets $X$ and $\sigma$  with a superpotential of the form 
\begin{equation}
W=X\left(\kappa\bar{\Psi}\Psi-M^2\right)+\beta\sigma\bar{\Psi}\Phi_{+}+\lambda S\Phi_{+}\Phi_{-},
\end{equation}
 where $M$ is some high energy scale, presumably the grand unified  scale. It is easy to show that the scalar components of the two-vector superfields    acquire vacuum expectation values $\langle\psi\rangle=\langle\bar{\psi}\rangle=M$, and $\langle X\rangle=\langle \sigma\rangle=0)$ which leave supersymmetry unbroken and $D$-term inflation unaffected. In this example, cosmic strings are  produced prior to the onset  of inflation and subsequently diluited. 

In conclusion, an inflationary stage dominated by a $D$-term avoids the slow-roll problem of inflation in supergravity and can naturally emerge in theories with a non-anomalous or anomalous $U(1)$ gauge symmetry. In  the latter  case, however, we have shown that   the scale of inflation as imposed by  the COBE  normalization is in contrast with the value  fixed by the Green-Schwarz mechanism of anomaly cancellation. We have  discussed      different possible solutions to this problem, {\it e.g.} the inclusion of a new term in the tree-level potential. We have also commented on the fact that at the end  of inflation global and local cosmic strings may be generated. If so, 
cosmic strings generate density perturbations which, in turn,  lowers the COBE normalized value of $\xi$ and exacerbate the problem of reconciling this value with the one  suggested by string theory. However, as we pointed out, this is  a very model-dependent issue and should be analyzed case by case.

\vskip 1cm
\underline{Acknowledgements}:

AR is grateful to G. Dvali for many enlightening discussions. The work of D.H. Lyth is partially supported by grants from PPARC and
from the European Commission
under the Human Capital and Mobility programme, contract
No.~CHRX-CT94-0423. A. Riotto is
supported by the DOE and NASA under Grant NAG5--2788.

\def\NPB#1#2#3{Nucl. Phys. {\bf B#1}, #3 (19#2)}
\def\PLB#1#2#3{Phys. Lett. {\bf B#1}, #3 (19#2) }
\def\PLBold#1#2#3{Phys. Lett. {\bf#1B} (19#2) #3}
\def\PRD#1#2#3{Phys. Rev. {\bf D#1}, #3 (19#2) }
\def\PRL#1#2#3{Phys. Rev. Lett. {\bf#1} (19#2) #3}
\def\PRT#1#2#3{Phys. Rep. {\bf#1} (19#2) #3}
\def\ARAA#1#2#3{Ann. Rev. Astron. Astrophys. {\bf#1} (19#2) #3}
\def\ARNP#1#2#3{Ann. Rev. Nucl. Part. Sci. {\bf#1} (19#2) #3}
\def\MPL#1#2#3{Mod. Phys. Lett. {\bf #1} (19#2) #3}
\def\ZPC#1#2#3{Zeit. f\"ur Physik {\bf C#1} (19#2) #3}
\def\APJ#1#2#3{Ap. J. {\bf #1} (19#2) #3}
\def\AP#1#2#3{{Ann. Phys. } {\bf #1} (19#2) #3}
\def\RMP#1#2#3{{Rev. Mod. Phys. } {\bf #1} (19#2) #3}
\def\CMP#1#2#3{{Comm. Math. Phys. } {\bf #1} (19#2) #3}

\end{document}